\documentclass[aps,prb,twocolumn,showpacs,floatfix,groupedaddress]{revtex4-1}

\usepackage[utf8]{inputenc}
\usepackage{amsmath}
\usepackage{amsfonts}
\usepackage{graphicx}
\usepackage{bm}
\usepackage{float}
\usepackage{mathtools}
\usepackage{verbatim}
\usepackage{flushend}

\begin{document}


\title{Probing the stability of the spin liquid phases in the Kitaev-Heisenberg model using tensor network algorithms}


\author{Juan Osorio Iregui}
\email[]{osorio@itp.phys.ethz.ch}
\affiliation{Theoretische Physik, ETH Zürich, 8093 Zürich, Switzerland.}

\author{Philippe Corboz}
\email[]{P.R.Corboz@uva.nl}
\affiliation{Theoretische Physik, ETH Zürich, 8093 Zürich, Switzerland.\\ Institute for Theoretical Physics, University of Amsterdam\\ 
			 Science park 904 Postbus 94485, 1090 GL Amsterdam, The Netherlands.}

\author{Matthias Troyer}
\email[]{troyer@phys.ethz.ch}
\affiliation{Theoretische Physik, ETH Zürich, 8093 Zürich, Switzerland.}


\date{\today}

\begin{abstract}
We study the extent of the spin liquid phases in the Kitaev-Heisenberg model using infinite Projected Entangled-Pair States tensor network ansatz wave functions directly in the thermodynamic limit. To assess the accuracy of the ansatz wave functions we perform benchmarks against exact results for the Kitaev model and find very good agreement for various observables. In the case of the Kitaev-Heisenberg model we confirm the existence of 6 different phases: Néel, stripy, ferromagnetic, zigzag and two spin liquid phases. We find finite extents for both spin liquid phases and discontinuous phase transitions connecting them to symmetry-broken phases.\\
\end{abstract}


\maketitle

\section{Introduction}

The field of frustrated magnets is arguably one of the most exciting areas of research in modern condensed matter physics. The competing interactions in these systems give rise to a rich variety of remarkable states of matter. In some cases, strong quantum fluctuations will preclude the formation of any conventional symmetry-breaking order, even in the limit of zero temperature, giving rise to so-called \emph{quantum spin liquids} (QSL).\cite{lee2008physics,mila2000quantum,balents2010spin} Interestingly enough certain instances of phases lacking a Landau-type description may give rise to so-called topological order,\cite{wen1990topological,wen1990ground} possibly hosting exotic anyonic excitations.\\

The possibility of finding novel materials exhibiting such exotic phases has motivated the search for QSLs from several directions. \cite{balents2010spin,meng2010quantum,coldea2001experimental} Among others, interesting proposals have been developed based on so-called Iridate compounds of the form $A_2IrO_3$ $(A=Na,Li)$. Even though these compounds are now known to be magnetically ordered at low temperatures \cite{liu2011long,singh2010antiferromagnetic,singh2012relevance} and the question of which effective model underlies their low-energy physics remains under debate with ideas ranging from spin-orbit coupled Mott insulators \cite{chaloupka2010kitaev,chaloupka2013zigzag} to the formation of quasi-molecular orbitals,\cite{mazin20122} there appears to be consensus on the fact that anisotropic exchange interactions play a key role. In particular, $Na_2IrO_3$ being an insulator with Curie-like magnetic susceptibility as well as a large spin-orbit splitting strongly suggests that it may belong to the category of spin-orbit coupled Mott insulators  with well-localized magnetic moments.\cite{chaloupka2013zigzag} These ingredients have sparked an interest in the possibility of finding anisotropic Kitaev-type physics \cite{kitaev2006anyons} in these and related compounds.\cite{kimchi2014kitaev,kimchi2013three} \\

However, the study of such highly correlated systems has posed tremendous challenges to their understanding as Monte Carlo based methods may fail to provide accurate results due to the infamous sign problem arising in generic fermionic or strongly frustrated spin models. \\

Over recent years the development of so-called \emph{tensor network algorithms} (TNA) opened the doors to a new approach towards understanding the physics of novel strongly correlated phases based on the notion of entanglement. One of the earliest triumphs of such entanglement-based algorithms was the development of the Density Matrix Renormalization Group \cite{white1992density} (DMRG) based on renormalization group ideas originally developed by Wilson.\cite{wilson1975renormalization} The DMRG is now understood to produce variational wave functions belonging to the class of so-called Matrix Product States (MPS) which have been rigorously shown to provide efficient approximations to the ground states of 1D gapped local hamiltonians.\cite{verstraete2006matrix} By now, two decades after its development, MPS (or the DMRG) have become the golden standard for the simulation of 1D lattice models. Furthermore, their 2D generalizations known as Projected Entangled-Pair States (PEPS)\cite{verstraete2004renormalization,verstraete2006criticality} (and their thermodynamic limit version infinite PEPS or iPEPS)\cite{jordan2008classical,corboz2010simulation} have been successfully used to study both fermionic systems as well as frustrated magnets,\cite{wang2013constructing,corboz2013tensor,corboz2012spin,matsuda2013magnetization,corboz2014crystals} with a notable example being the lowest variational energies for the t-J model available to date for large systems.\cite{corboz2014competing} \\

In what follows we will make use of the iPEPS ansatz to study the simple yet remarkably rich \emph{Kitaev-Heisenberg} model \cite{chaloupka2010kitaev,chaloupka2013zigzag} originally developed to encode the low energy degrees of freedom in the iridates. In particular, we will focus on the stability of the spin liquid phases around the Kitaev points in the model and we will provide strong evidence for their survival over a considerable extent of the phase diagram in the thermodynamic limit.

\section{The Kitaev-Heisenberg model}
A very interesting proposal was recently put forward by Chaloupka et al.\cite{chaloupka2013zigzag} based on the \emph{Kitaev-Heisenberg} model\cite{chaloupka2010kitaev} yet extended to its full parameter space, i.e.
\begin{equation}
	H_{i,j}^{(\gamma)} = A \left( \cos \varphi  \; \vec{S}_i \cdot \vec{S}_j + 2 \sin \varphi \; S_i^{(\gamma)} S_j^{(\gamma)} \right), 
	\label{eq:heis_kit_ham}
 \end{equation}

with $\left(i,j\right)$ labeling nearest-neighbour sites of a honeycomb lattice, the first term being an isotropic Heisenberg interaction, the second an anisotropic Kitaev interaction in which $\gamma \in (x,y,z)$ determines the spin components interacting along a given bond, $\varphi \in [-\pi,\pi)$ and $A$ an overall scaling factor which we set to one. \\

This model has been tackled in the past using a variety of approaches in different regions of its full parameter regime.\cite{schaffer2012quantum,jiang2011possible,chaloupka2010kitaev,chaloupka2013zigzag,he2013study} In its original formulation \cite{chaloupka2010kitaev} (covering only the region $\varphi\in[-\pi/2,0]$) small system studies \cite{chaloupka2010kitaev,jiang2011possible} found either a second or weak first  order phase transition joining a spin liquid phase to a so-called stripy phase at roughly $\varphi\approx-76^{\circ}$ (or $\alpha\approx0.8$ in the original formulation), see Fig.~\ref{fig:order_patterns}. This value is also reported in the extended formulation on a 24-site system.\cite{chaloupka2013zigzag} The restricted formulation was also studied from a slave-particle mean field approach \cite{schaffer2012quantum} where the behaviour of the order parameter itself showed a discontinuity, yet it was suggested that the transition could indeed be of either second or weak first order type at a value of $\varphi\approx-72^{\circ}$ ($\alpha\approx0.76$), owing to the lack of quantum fluctuations beyond the mean field level in their approach. Finally, a so-called mixed PEPS (mPEPS) approach found a phase transition at $\varphi\approx-89^{\circ}$ ($\alpha \approx 0.99$).\cite{he2013study} \\ 

In this study we approach the model directly in the thermodynamic limit and provide results that go systematically beyond the mean field level using iPEPS.\footnote{Here, however, we do not claim that the iPEPS mean field ansatz provides a quality equivalent to the slave-particle mean field study in Ref.~\onlinecite{schaffer2012quantum} which was designed to represent the kitaev point exactly.}

\section{The Method (iPEPS)}
\begin{figure}[H]

 \centering
 \begin{tabular}{c}
 \includegraphics[width=0.31\textwidth]{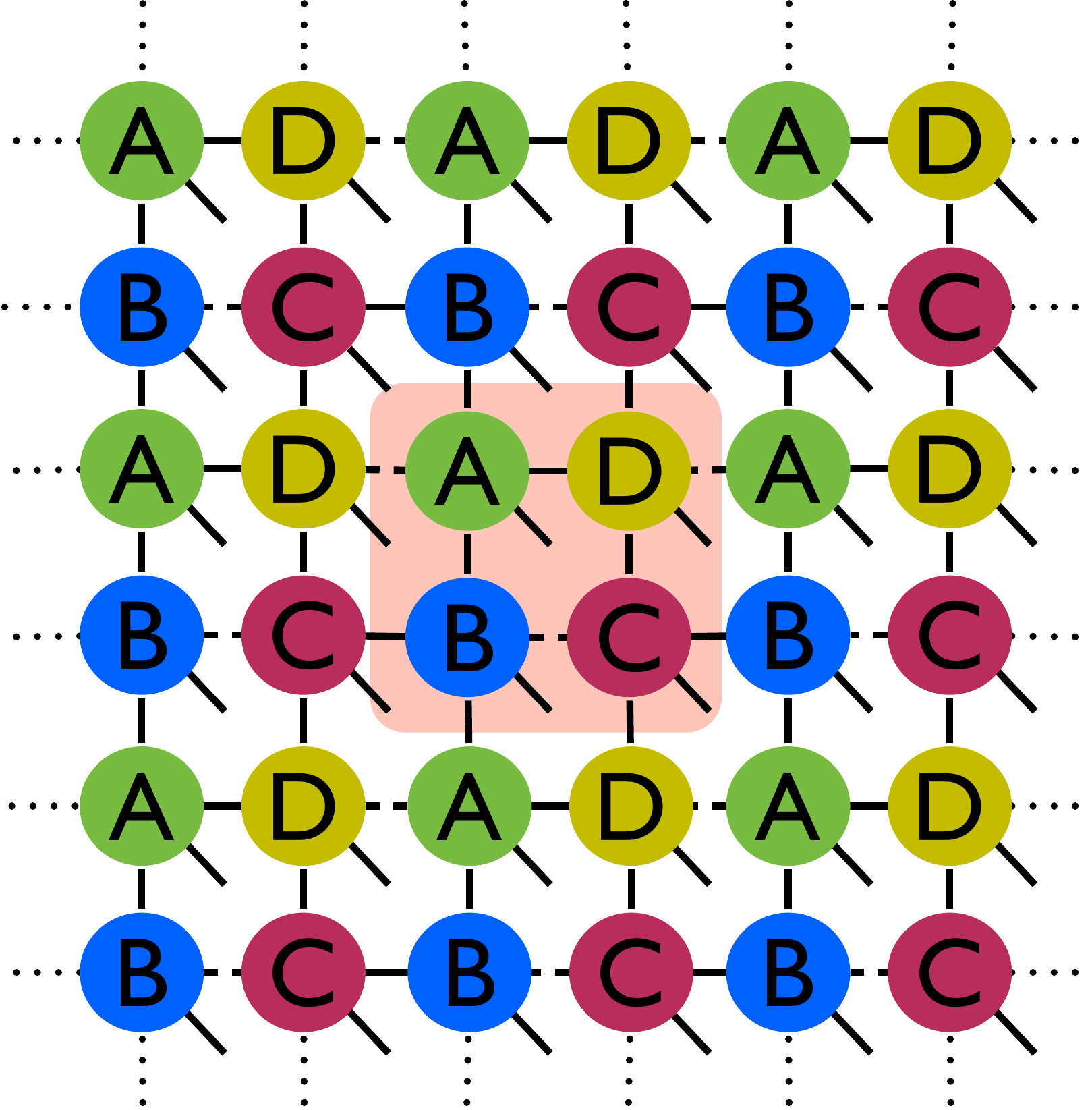} \\
 \includegraphics[width=0.49\textwidth]{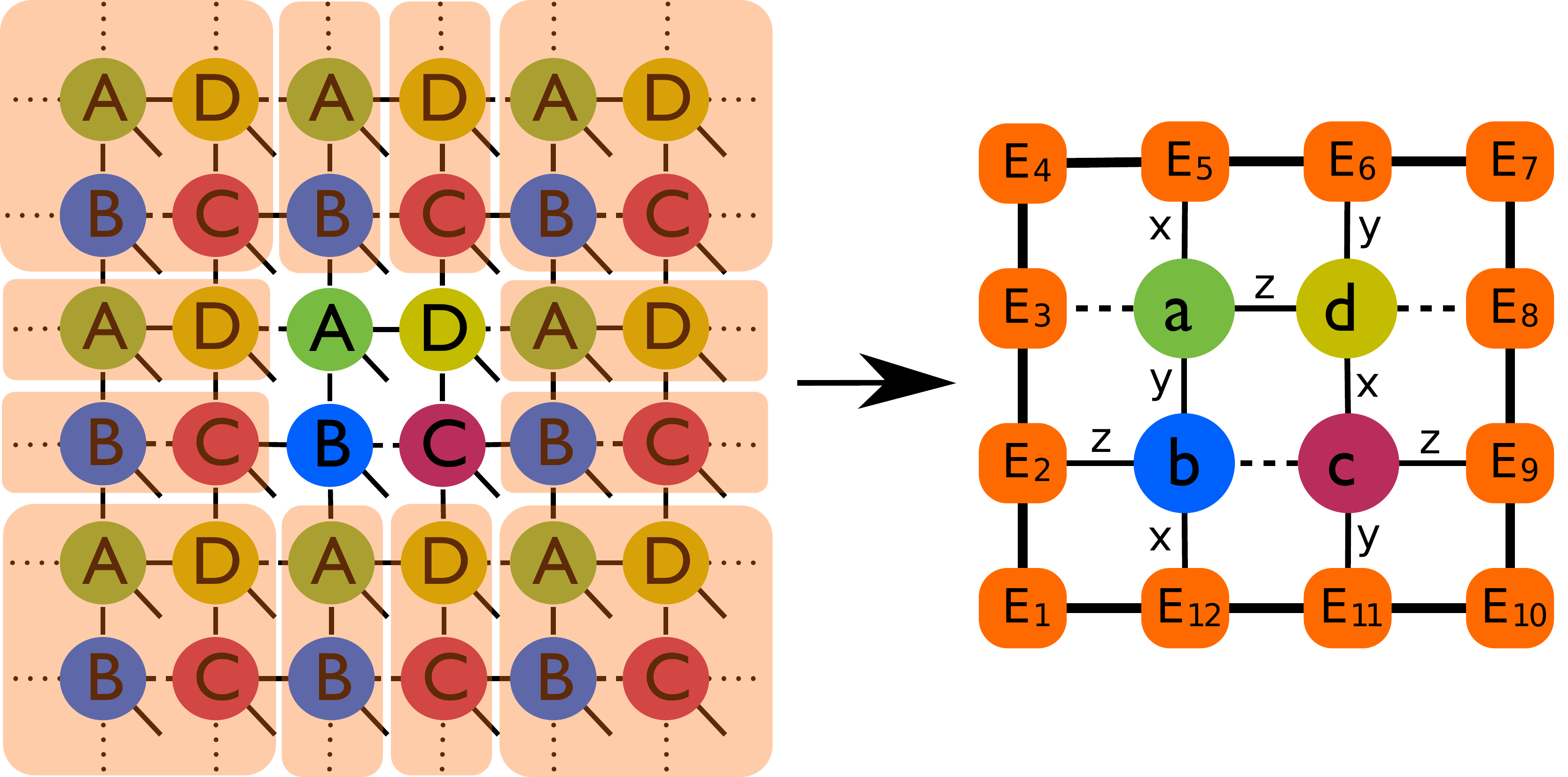}
 \end{tabular}	
 
\caption{\label{fig:iPEPS} Top: 4-site unit cell iPEPS ansatz on the brick-wall lattice. Introducing trivial (dashed) bonds in a square lattice leads to a so-called brick-wall lattice reproducing the connectivity of the honeycomb lattice. The blue, red, green and purple tensors (circles) in the unit cell (highlighted cluster) provide the variational parameters. Bottom: Effective mapping used for the evaluation of observables. On the left the ket portion of the network prior to contraction. On the right a contracted bra-ket network is shown from the top with tensors labeled by lowercase letters representing a bra-ket pair contracted. The orange tensors (squares) labelled by \emph{E} represent environment tensors, obtained eg. via the CTM algorithm, accounting for an effectively infinite environment. Thick lines (bonds) correspond to indices of dimension $\chi$, solid lines to indices of dimension $D$ and dashed lines represent trivial bonds. Bond labeling convention adopted is given by the x,y,z labels.}
\end{figure}

In general, TNA rely on the use of tensors as variational parameters used for the optimization of the ansatz wave function, i.e. in the TNA approach ansatz wave functions take on the form

\begin{equation}
	\lvert \psi \rangle = \sum_{\{S_r\}_r} Tr \prod_{\bf r} A[{\bf r}]_{S_r} \lvert \{ S_r\}_r\rangle, 
	\label{eq:tna_wavefunction}
 \end{equation}

where $A[{\bf r}]$ are tensors of rank $k+1$, with $k$ (typically) the coordination number of the lattice and an additional index labelling the local basis elements, with $\lvert\{ S_r\}_r\rangle = \lvert S_{{\bf r_1}},S_{{\bf r_2}},...,S_{{\bf r_N}}\rangle$ a many-body basis state (${\bf r_i}$ labeling the different lattice sites) and $Tr$ symbolizing the contraction of all auxiliary (also called virtual) indices between adjacent tensors, see Fig.~\ref{fig:iPEPS}. 

The dimension of the virtual indices is refered to as the \emph{bond dimension} and is usually denoted by $D$. The bond dimension determines the amount of entanglement that can be encoded in the wave function and controls the underlying accuracy of the algorithm. At the heart of the efficiency of the algorithms lies the fact that the number of variational parameters only scales polynomially with $D$ and linearly with the system size for finite systems.

\subsection{Ansatz}
In the present study we formulate the iPEPS ansatz on the honeycomb lattice by mapping it onto a brick-wall lattice in which the connectivity of the honeycomb lattice is reproduced exactly by introducing a trivial index on each tensor.\cite{corboz2012spin} We have studied wave functions made of up to $2$ by $2$ unit cells with all four (rank 5) tensors being independent, see Fig.~\ref{fig:iPEPS}. Despite the increase in computational complexity tensors with complex entries were employed as these yielded the best results.

\subsection{Contraction}
As mentioned above, in the TNA approach an efficient representation of ground state wavefunctions can be achieved via a tensor decomposition ansatz. It should be noted, however, that in the case of PEPS the computational complexity of contracting a full tensor network involved in eg. the evaluation of expectation values of observables increases exponentially with system size and thus can only be evaluated in an approximate way.\\

In practice approximate contraction schemes based on corner transfer matrices (CTM),\cite{nishino1996corner,orus2009simulation} MPS,\cite{cirac2011entanglement,lubasch2014unifying} tensor-entanglement renormalization group (TERG)\cite{levin2007tensor,gu2008tensor} and higher-order (HOTRG) variants\cite{xie2012coarse,yu2014tensor} are used for the contraction of tensor networks both in the finite and infinite cases. Here we have relied on the directional CTM scheme of Ref.~\onlinecite{orus2009simulation} for constructing the tensor networks involved in both the optimization of the tensors as well as the evaluation of observables. In this scheme the contraction of the infinite bra and ket tensor networks surrounding a unit cell is effectively represented by introducing a boundary made up of so-called environment tensors, see Fig.~\ref{fig:iPEPS}. The environment tensors are constructed via iterative absorption and renormalization of unit cell tensors into the boundary tensors in Fig.~\ref{fig:iPEPS}. Importantly, the accuracy of the contraction is controlled by the bond dimension of the environment tensors, usually denoted by $\chi$. For the data presented here $\chi$ was chosen to be larger than $D^2$ in all cases and large enough to yield negligible variations in the energies.  For a more precise description of the details involved in the contraction scheme we refer the reader to Ref.~\onlinecite{orus2009simulation}.

\subsection{Optimization}
Optimization of the tensors generating the ansatz wave functions is typically performed using either direct energy minimization or imaginary-time evolution. Here we have used the latter combined with the so-called full update scheme.\cite{corboz2010simulation} In the imaginary-time evolution procedure by starting from some initial state $\lvert\psi_0\rangle$ of the form (\ref{eq:tna_wavefunction}) we perform subsequent projection steps

\begin{equation}
	\lvert \psi_{k+1} \rangle = \frac{\mathrm{e}^{-\tau \hat{H}} \lvert\psi_k\rangle}{\| \mathrm{e}^{-\tau \hat{H}} \lvert \psi_k \rangle \|}, 
	\label{eq:imaginary_time}
 \end{equation}

so provided that the initial state $|\psi_0\rangle$ had some overlap with the ground state of the model enough iterations will eventually converge to the ground state.\\

For the data presented it was observed that values below $\tau = 0.01$ for the imaginary-time evolution did not provide a significant improvement in the quality of the data. In all cases the number of cumulative iterations was such that it lead to values of at least $ \beta = 20$ and in all cases it was found to be large enough to achieve convergence of the variational energies. Here we point out that lower cost variants like the simple update,\cite{jiang2008accurate} in which an explicit construction of the environment is omited, failed to yield good results in the Kitaev limit and thus we opted for performing all the simulations using the full update, inspite of its significantly larger computational cost.\cite{corboz2010simulation}

\section{Kitaev Limit Benchmarks} 
In the limits $\varphi = \pm 90^{\circ}$ the model in Eq. (\ref{eq:heis_kit_ham}) becomes the well-known Kitaev honeycomb model \cite{kitaev2006anyons} with equal bond couplings (B-phase). Indeed, even though the interactions on each bond are of Ising type, the fact that different bonds correspond to different quantization axes makes the Kitaev model a highly frustrated one, even classically, since it is impossible to satisfy all energy constraints simultaneously. From the exact solution of the Kitaev model \cite{kitaev2006anyons} it is known that inside the B-phase two different types of excitations arise: magnetic vortices which are gapped and localized in the absence of an external magnetic field and gapless Majorana fermions moving in the static background field of the vortices. Perhaps more interestingly, these excitations can be gapped into a topological phase exhibiting non-abelian anyonic statistics.\cite{kitaev2006anyons}\\

\begin{figure}[H]
\includegraphics[width=0.54\textwidth]{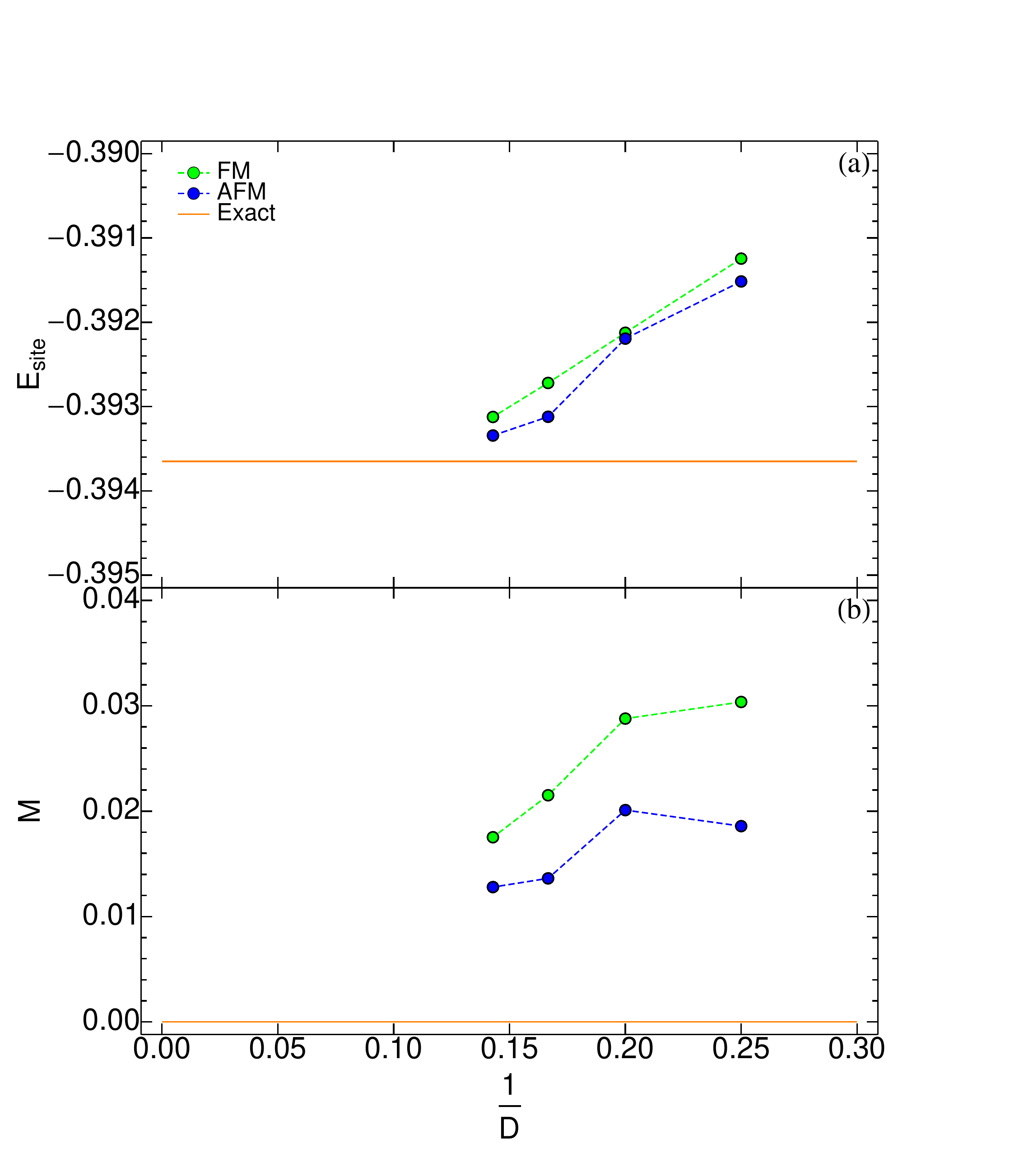}
\caption{\label{fig:energy_mag} (a) Energy per site and (b) Magnetization as a function of inverse bond dimension. Errors in the energy for the largest value studied ($D = 7$) are of the order of $10^{-4}$. Magnetization values are normalized to 1.}
\end{figure}

From both Kitaev's seminal paper \cite{kitaev2006anyons} as well as later work \cite{baskaran2007exact} it can be gathered that the energy per site for this model at the equal coupling limit considered here is $E_{site} = -0.3936$ independent of the nature of the couplings, i.e. for both ferromagnetic as well as antiferromagnetic couplings. Our best variational approximations to the energy per site are $E_{site}^{FM} = -0.3931$ and $E_{site}^{AFM} = -0.3933$ with a bond dimension $D=7$ ($\chi = 60$), yielding good agreement with the exact value, see Fig.~\ref{fig:energy_mag}.\\

A feature of the Kitaev model is that the ground state is known to be a $\mathbb{Z}_2$ spin liquid and as such develops no local order parameter. In our case, we find variational states exhibiting a strongly suppressed magnetization, with the largest values of the magnetization being around 0.03 and 0.02 in the ferromagnetic and antiferromagnetic cases, respectively, with a bond dimension $D=4$. The level of symmetry breaking is observed to decrease in general as a function of increasing $D$ (entanglement) and for our best variational states the magnetization reaches a minimum of approximately 0.02 for the ferromagnetic case and 0.01 for the antiferromagnetic case, with $D=7$. See Eqs. (\ref{eq:mag})-(\ref{eq:neel}) below for our definition of magnetization as well as similar magnetic order parameters.\\

The fact that the model is strongly frustrated and exhibits gapless excitations turns it into a formidable challenge for numerical methods in general. In what follows we will show that iPEPS ansatz wave functions are capable of capturing the essential features of this model quite well, even in the absence of an energy gap, using only a modest value of the bond dimension $D$.\\

As mentioned above the bond dimension $D$ controls the amount of entanglement in the wave functions. This means that one may systematically tune the degree to which quantum fluctuations manifest themselves in the ansatz state. In particular, the case $D=1$ represents a product state and as such leads to a mean field level ansatz potentially overestimating the degree of symmetry breaking inside a phase. Upon increasing $D$ additional quantum fluctuations allow for a renormalization of observables such as the magnetization and thus lead to the obeserved suppression above.

\begin{figure}[H]
\includegraphics[width=0.53\textwidth]{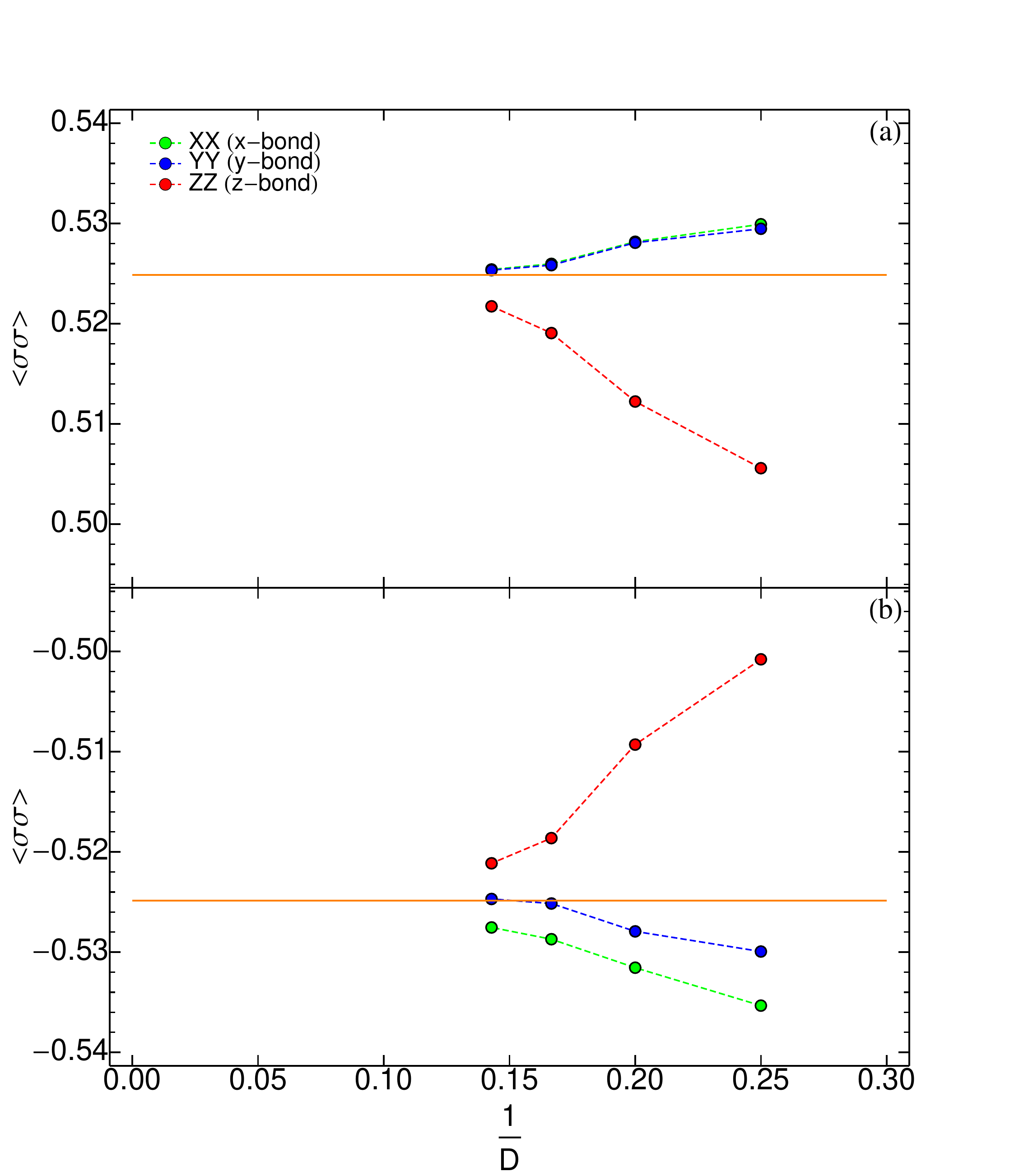}
\caption{\label{fig:correlators} Nearest neighbor correlators for corresponding bond types in the ferromagnetically coupled case (a), and antiferromagnetically coupled case (b). All remaining correlators are at most of the order of $10^{-3}$ for the largest value of $D$ and exhibit similar improvement with increasing $D$.}
\end{figure}

Another remarkable feature of the Kitaev model has to do with the peculiar form of the spin-spin correlators, for which it is known that only spin components matching the type of a given bond will exhibit non-vanishing correlations \cite{baskaran2007exact} i.e. $ \left< \sigma_i^{(\gamma)} \sigma_j^{(\gamma)} \right> = 0.525 $ iff $ \gamma = \left( i , j \right) $. Examining the spin correlators we find that the correlator structure is well represented by our ansatz wave functions with all correlators heading towards the exact values monotonically as the bond dimension increases, see Fig.~\ref{fig:correlators}. Here a curious stronger deviation for the Z-type correlators can be initially observed. This is nothing but a consequence of the way we map the honeycomb lattice onto a brick-wall lattice when constructing the environment. Importantly, this feature is systematically reduced as we increase the amount of variational parameters nicely reflecting the putative non-symmetry broken nature of our wave functions.\\

Indeed, the quality of our data could be improved even further by increasing the value of the bond dimension $D$. We have, however, not pushed our simulations beyond this point as we believe the current data already provide strong support for the spin liquid character of our variational wave functions.
 
\section{Kitaev-Heisenberg Results}
\begin{figure}[H]
 \centerline{\includegraphics[scale=0.55]{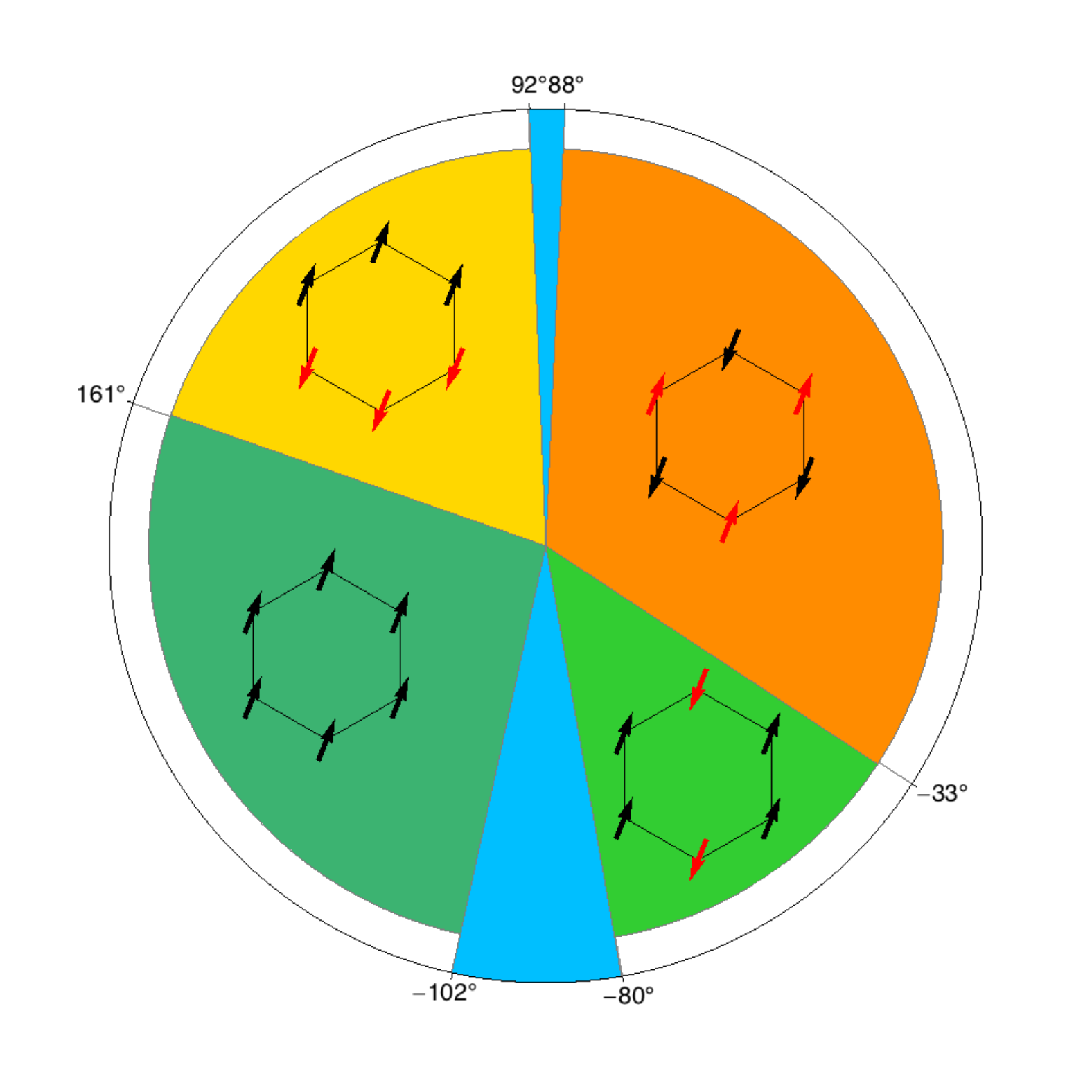}}
 \caption{\label{fig:order_patterns} Regions spanned by the different phases found using iPEPS. Four different magnetically ordered (collinear) phases are found: Néel (orange/top right), zigzag (yellow/top left), ferromagnetic (dark green/bottom left) and stripy (light green/bottom right). Magnetically ordered phases are characterized using the order parameters in (\ref{eq:mag})-(\ref{eq:neel}). Blue regions correspond to the spin liquid phases. Phase boundary angles correspond to $D=6$ estimates with the spin liquid regions increasing their extent as the bond dimension $D$ is increased.}
 \end{figure}

\begin{figure}[H]
 \begin{tabular}{l}
 \includegraphics[width=0.50\textwidth]{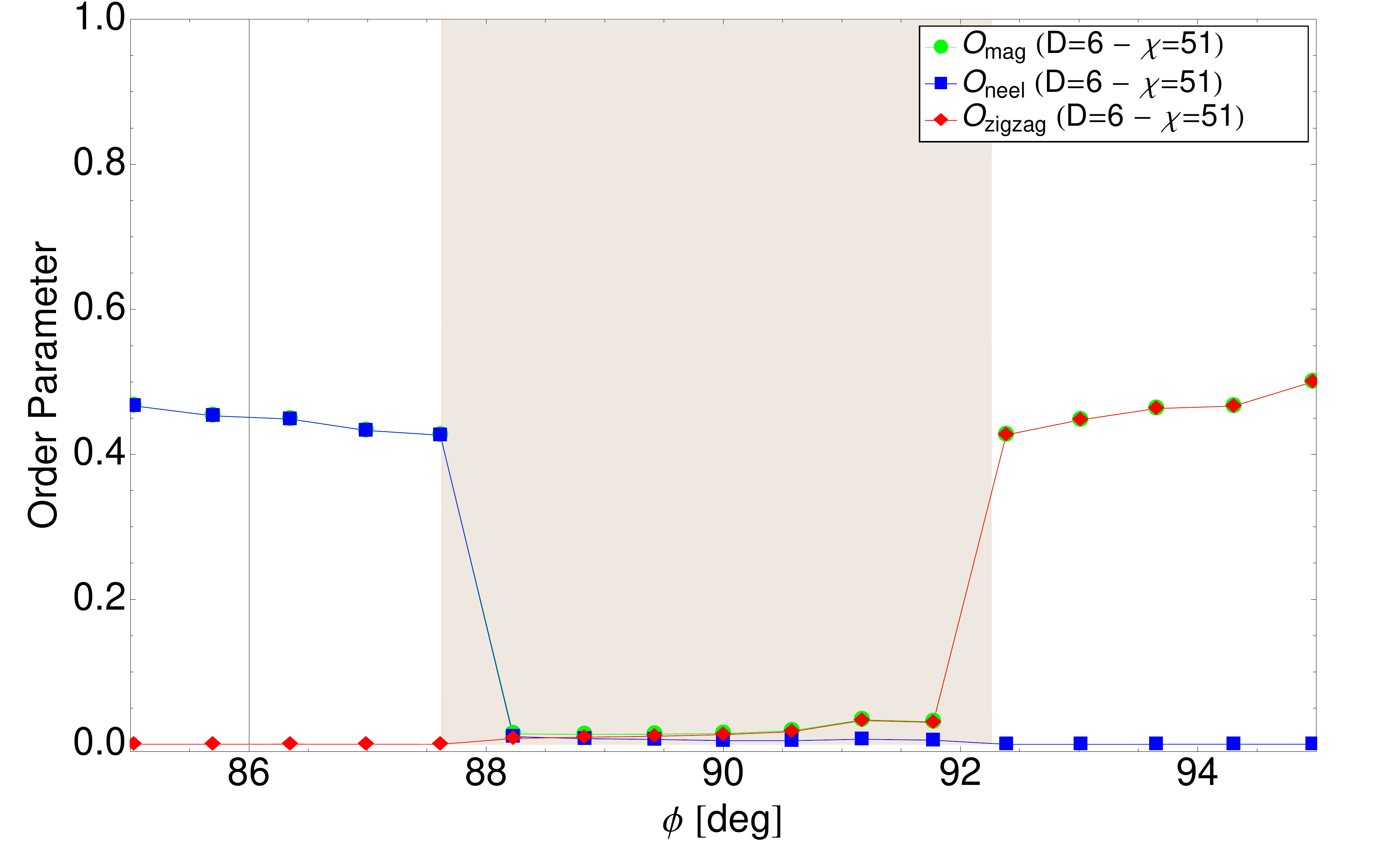} \\
 \includegraphics[width=0.50\textwidth]{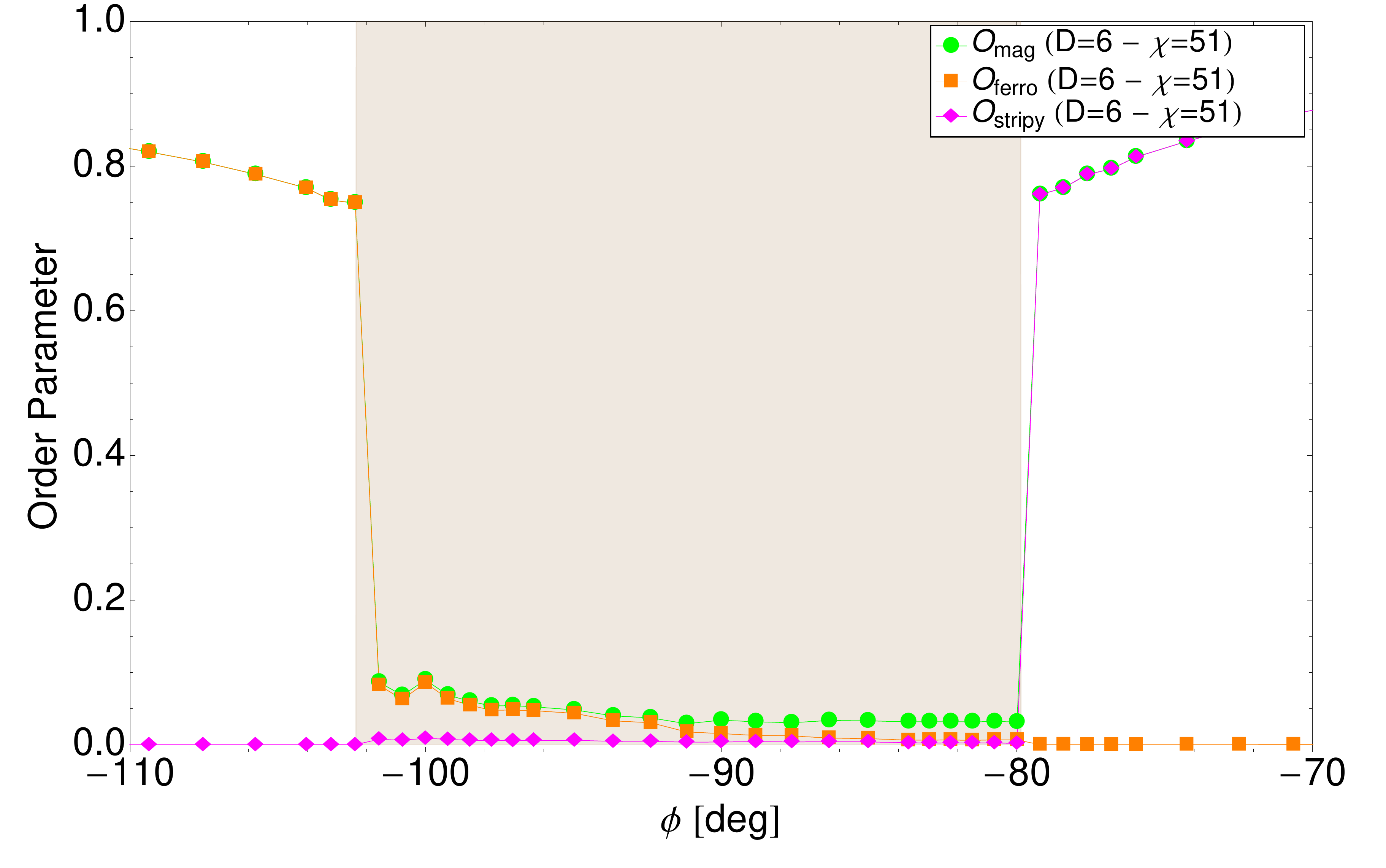}
 \end{tabular}
 \caption{ \label{fig:order_parameters} Order parameters as a function of the angle $\varphi$ in the vicinity of the points $\varphi = \pm 90^{\circ}$, normalized to 1. Regions of strongly suppressed symmetry breaking are clearly visible in both cases with finite discontinuities in the order parameters. The shaded regions indicate the estimated extents of the regions covered by the QSL phases. All order parameters not shown remain very close to zero.}
 \end{figure}

Having established the accuracy of our ansatz wave functions in the Kitaev limit we proceed to evaluate the stability of the spin liquid phases in the Kitaev-Heisenberg model of Eq. (\ref{eq:heis_kit_ham}). In their proposal Chaloupka et al. \cite{chaloupka2013zigzag} performed a 24-site Lanczos diagonalization study of this model in which 6 different phases were identified, namely: antiferromagnetically coupled QSL (ASL), ferromagnetically coupled QSL (FSL), Néel, stripy, ferromagnetic and zigzag. We have found the same phases using iPEPS, see Figs. \ref{fig:order_patterns} and \ref{fig:order_parameters}.

  In the study by Chaloupka et al. the phase transitions between symmetry broken phases (stripy$/$Néel and ferromagnetic$/$zigzag) were found to be of first order, whereas the FSL to ordered transitions were found to be of either second or weak first order.\cite{chaloupka2013zigzag}  In the case of the ASL phase, the nature of the transitions to the ordered zigzag and Néel phases was not directly identified but observed to correspond to level crossings, thus pointing towards first order-type transitions. For a summary of the locations of the phase transitions found in the small cluster, see Table \ref{tab:transitions}.\\
 
Here, in order to capture the different types of magnetic order we define 5 different order parameters, i.e.
\begin{align}
	O_{mag}& = \sqrt{ \frac{1}{4} \left( \langle\vec{\sigma}_A\rangle^2 + \langle\vec{\sigma}_B\rangle^2 + \langle\vec{\sigma}_C\rangle^2 + \langle\vec{\sigma}_D\rangle^2 \right) }, \label{eq:mag} \\
	O_{ferro}& = \sqrt{ \frac{1}{4} \left(\langle \vec{\sigma}_A\rangle + \langle\vec{\sigma}_B\rangle + \langle\vec{\sigma}_C\rangle + \langle\vec{\sigma}_D\rangle \right)^2 }, \label{eq:ferro} \\
	O_{stripy}& = \sqrt{ \frac{1}{4} \left(\langle \vec{\sigma}_A\rangle - \langle\vec{\sigma}_B\rangle - \langle\vec{\sigma}_C\rangle + \langle\vec{\sigma}_D\rangle \right)^2 }, \label{eq:stripy} \\
	O_{zigzag}& = \sqrt{ \frac{1}{4} \left(\langle \vec{\sigma}_A\rangle + \langle\vec{\sigma}_B\rangle - \langle\vec{\sigma}_C\rangle - \langle\vec{\sigma}_D \rangle \right)^2 }, \label{eq:zigzag} \\
	O_{\text{\emph{Néel}}}& = \sqrt{ \frac{1}{4} \left(\langle \vec{\sigma}_A\rangle - \langle\vec{\sigma}_B\rangle + \langle\vec{\sigma}_C\rangle - \langle\vec{\sigma}_D \rangle \right)^2 }, \label{eq:neel}
\end{align}
with the first one providing a signature for \emph{any} form of magnetic order and the rest being designed to identify the different types of order expected inside different regions of the phase diagram.\\ 

 To locate different phase transitions, we proceed as follows. Starting from a set of preliminary runs across the phase diagram in which the different competing phases are identified by performing imaginary time evolution for all angles $\varphi\in[-\pi,\pi)$ we locate regions far away from potential, yet to be accurately determined, phase transitions. Once we have optimized tensors well inside each phase we use these tensors as starting states for a second set of imaginary-time evolution runs across a given range of $\varphi$ values and compare the energies of the two (competing) phases inside this angle window.\\ 
 
 \begin{table}[H]
 \centering
 \begin{tabular}{l|c|c|}
 	\cline{2-3}
 	 & \multicolumn{1}{ c| }{iPEPS} & \multicolumn{1}{ c| }{Lanczos} \\
 	\hline
 	\multicolumn{1}{|c|}{ASL - Néel} & $88^{\circ}$ & $88^{\circ}$ \\ \hline
 	\multicolumn{1}{|c|}{ASL - Zigzag} & $92^{\circ}$ & $92^{\circ}$ \\ \hline
 	\multicolumn{1}{|c|}{FSL - Stripy} & $-80^{\circ}$ & $-76^{\circ}$ \\ \hline
 	\multicolumn{1}{|c|}{FSL - Ferro} & $-102^{\circ}$ & $-108^{\circ}$ \\ \hline
 	\multicolumn{1}{|c|}{Ferro - Zigzag} & $161^{\circ}$ & $162^{\circ}$ \\ \hline
 	\multicolumn{1}{|c|}{Stripy - Néel} & $-33^{\circ}$ & $-34^{\circ}$ \\ \hline
 \end{tabular}
 \caption{ \label{tab:transitions} Transition points between different phases found using iPEPS with a bond dimension $D=6$. We include the original 24-site Lanczos results from Chaloupka et al. \cite{chaloupka2013zigzag} for reference.}
 \end{table}
 
 The reasoning behind this procedure relies on the fact that in the presence of a first order phase transition we expect to observe a certain amount of hysteresis as we cross the transition point. This should allow us to evaluate the energies of two different (competing) phases across a (possibly narrow) parameter range, thus making a potential level crossing visible. This, together with the behavior of the different order parameters as the transition is crossed, should allow us to identify the type of the transition.
 
\subsection{FSL-Stripy transition}
\begin{figure}[H]
 \begin{tabular}{l}
 \includegraphics[width=0.484\textwidth]{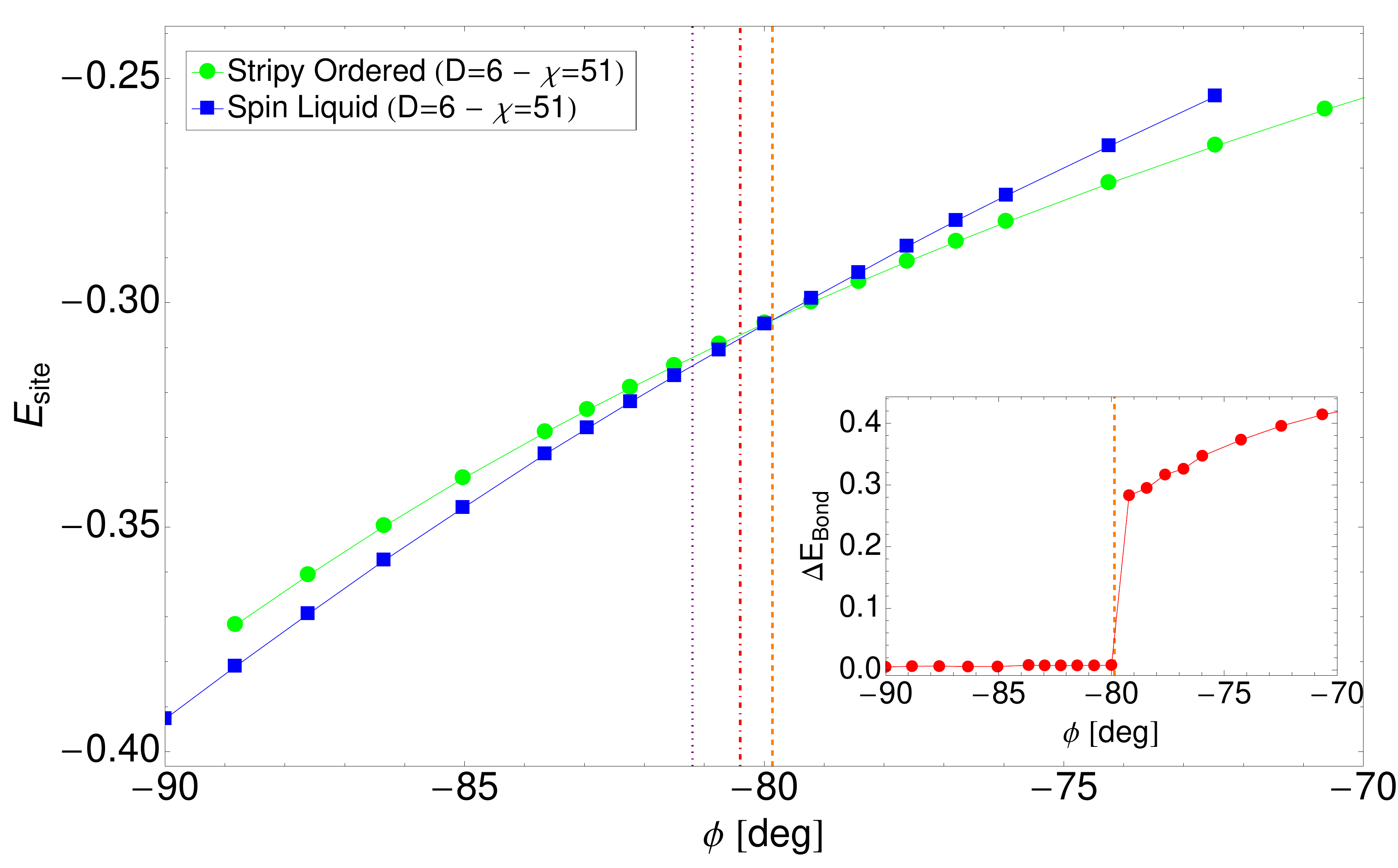} \\
 \,\,\, \includegraphics[width=0.48\textwidth]{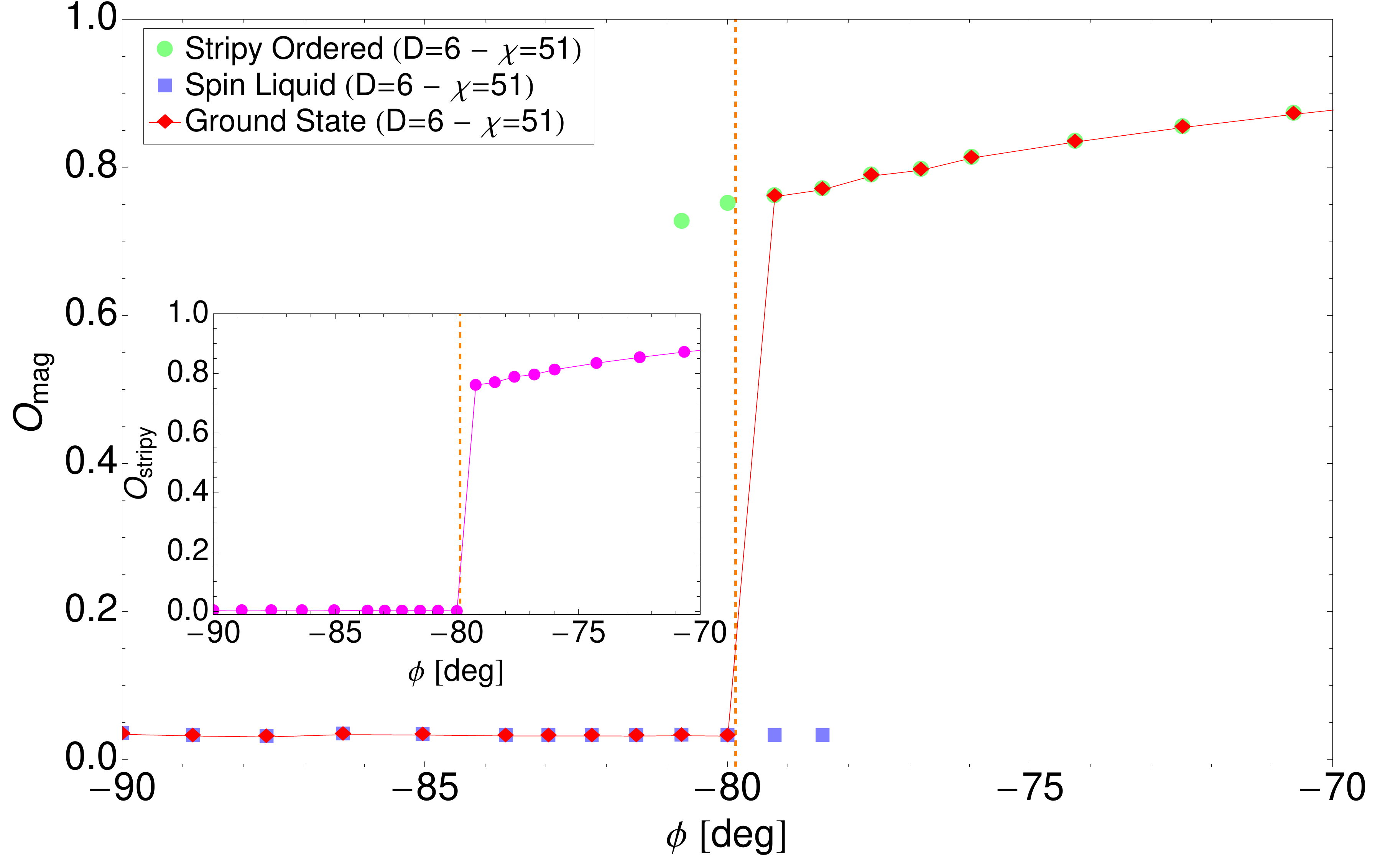}
 \end{tabular}
 \caption{ \label{fig:crossing_analysis_FQSL_stripy_1} Top: Energy crossings for the FSL to stripy phase transition. Dotted (Purple), Dash-Dotted (Red) and Dashed (Orange) lines correspond to the location of the phase transition for $D=4,5,6$, respectively. The inset shows the maximum deviation of the bond energies from the mean value normalized to the mean value.   Bottom: $O_{mag}$ within each phase (green circles: stripy / blue squares: FSL) and the reconstructed ground state curve (red diamonds). The inset shows the behaviour of $O_{stripy}$ over the same range of angles. Order parameters are normalized to one. Dashed orange lines indicate the estimated location of the phase transition for $D = 6$.}
 \end{figure}

In order to illustrate the procedure proposed above let us consider the region corresponding to FM Kitaev couplings and AFM Heisenberg couplings defined by $\varphi \in [-90^{\circ},0]$. After a preliminary set of runs we find 3 different phases in this quadrant: FSL, stripy and Néel as one moves from $\varphi = -90^{\circ}$ to $\varphi=0$. Noting that this matches the results from previous studies \cite{schaffer2012quantum,jiang2011possible,chaloupka2010kitaev,chaloupka2013zigzag,he2013study} it remains to verify how our results agree with the transition points found previously. Having obtained tensors representing the phases at the two extremes of the angle window shown in Fig.~\ref{fig:crossing_analysis_FQSL_stripy_1} we perform imaginary-time evolution on these states for all values of $\varphi$ in this window.\\

There an energy crossing at a finite angle is clearly observed for a value of $\varphi \approx -80^{\circ}$ with $D=6$. We argue that this is an actual level crossing in the system. Moreover, the magnetization data shows a pronounced jump consistent with a first order phase transition, see bottom plot in Fig.~\ref{fig:crossing_analysis_FQSL_stripy_1}. Here all order parameters remain remarkably close to zero within the $\varphi \in [-90^{\circ},-80^{\circ}]$ range, as expected for the FSL, and a jump in both $O_{mag}$ and $O_{stripy}$ occurs as the energies of the FSL and stripy phases cross, indicating a transition into a stripy ordered phase. In the upper inset in Fig.~\ref{fig:crossing_analysis_FQSL_stripy_1} we show the relative deviation in bond energies from the average value and here a jump is also visible. Similarly, the transition between symmetry-broken stripy and Néel phases is found to be of first order and located at $\varphi \approx -33^{\circ}$ (data not shown).
 
 We also note that the phase boundary here systematically shifts towards smaller values of $\varphi$ (in norm) as we increase the bond dimension $D$, effectively increasing the size of the FSL region. A linear extrapolation in $1/D$ to the $D\rightarrow\infty$ limit yields a phase boundary at $\varphi_{\infty} \approx -77^{\circ} $, very close to the value found in previous studies\cite{jiang2011possible,chaloupka2010kitaev,chaloupka2013zigzag} i.e. $\varphi \approx -76^{\circ}$ (or $\alpha \approx 0.8$ in the original parametrization\cite{chaloupka2010kitaev}). As it is not clear that such an extrapolation should hold we quote the transition values for $D = 6$ in Table \ref{tab:transitions} and interpret this value as a \emph{lower bound} for the  extent of the FSL phase on this part of the phase diagram. 

 \begin{figure*}
 \centering
 \begin{tabular}{ccc}
 \includegraphics[width=0.311\textwidth]{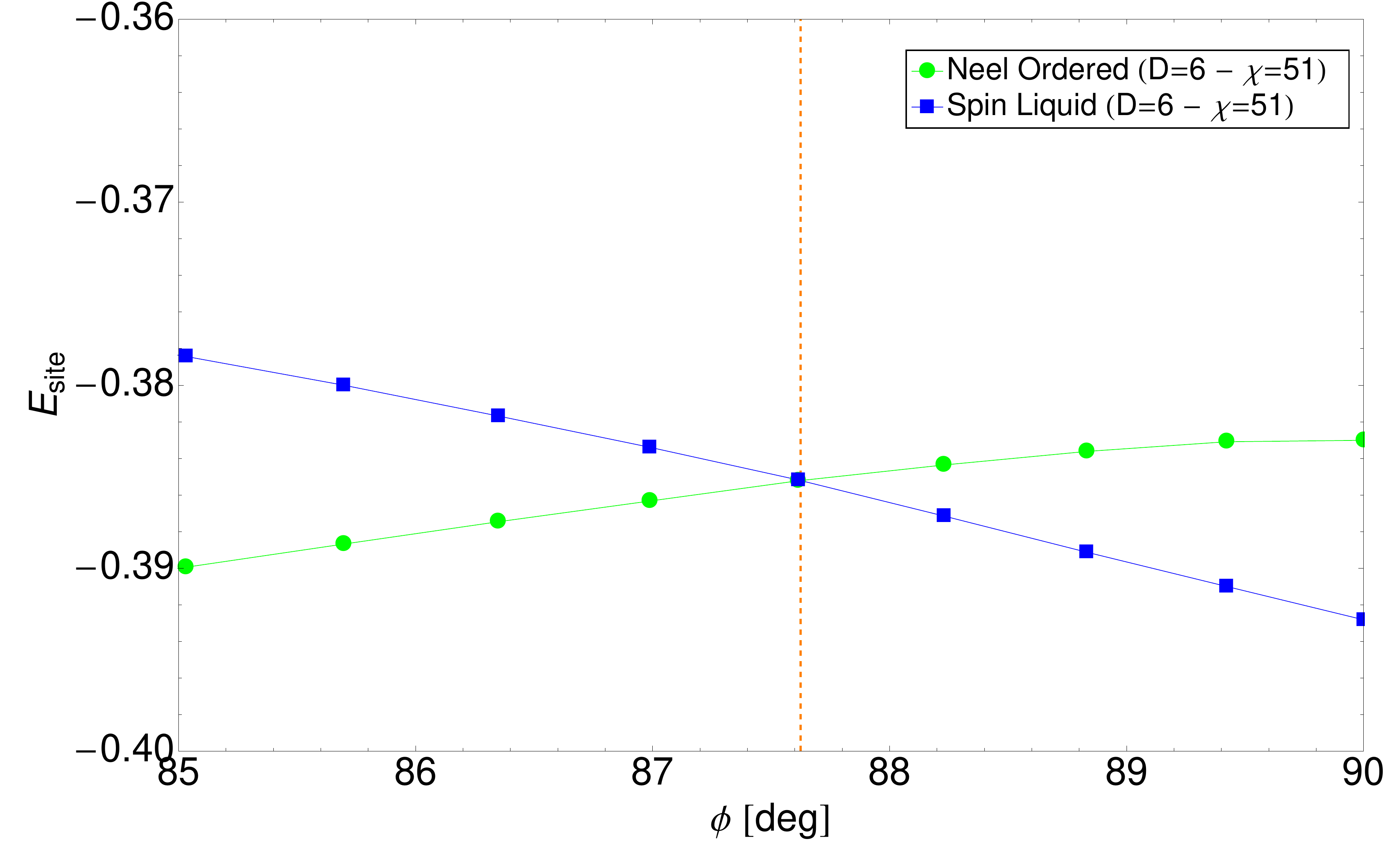} & \includegraphics[width=0.31\textwidth]{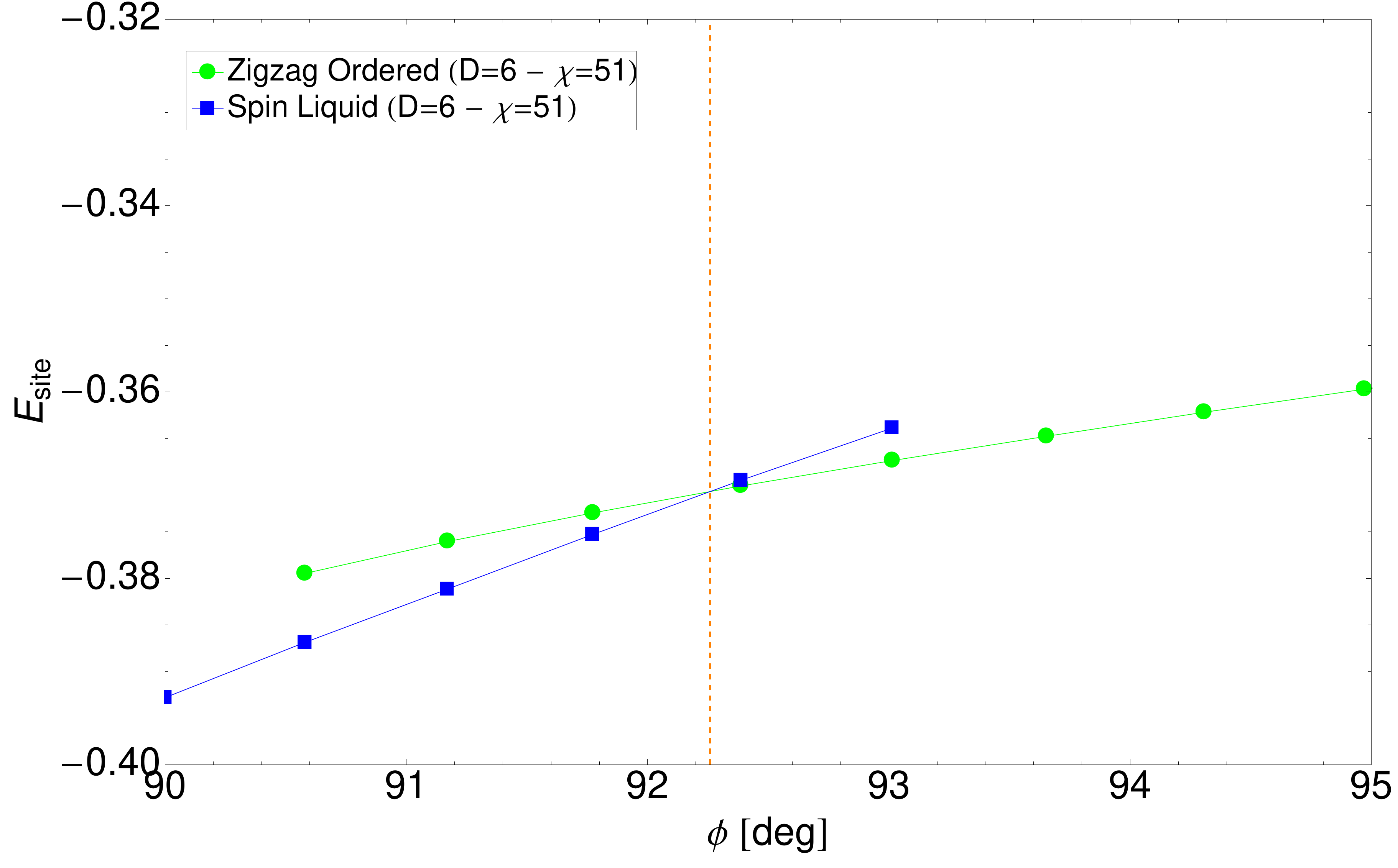} & \includegraphics[width=0.312\textwidth]{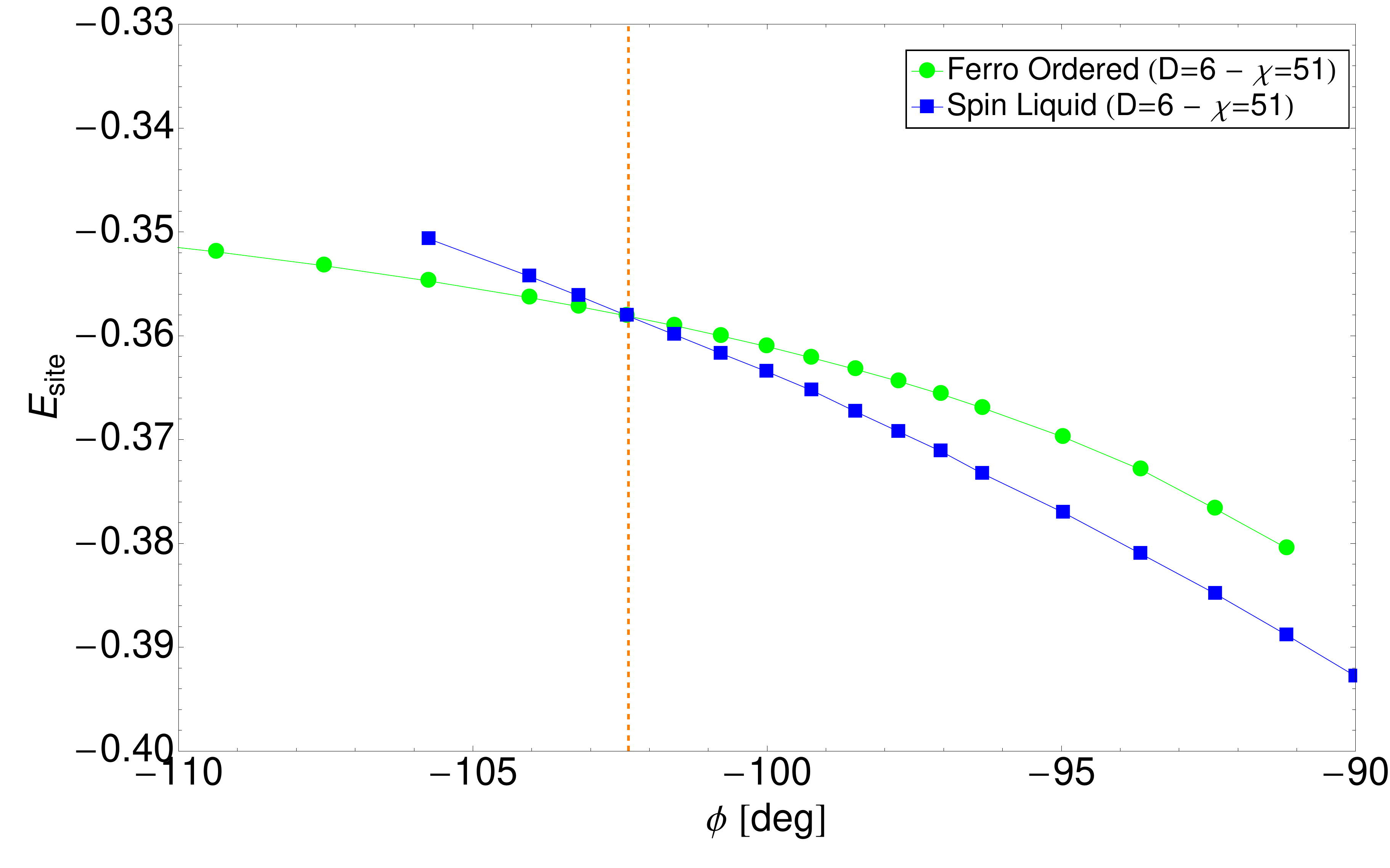} \\
 \,\,\,\,\,\,\includegraphics[width=0.312\textwidth]{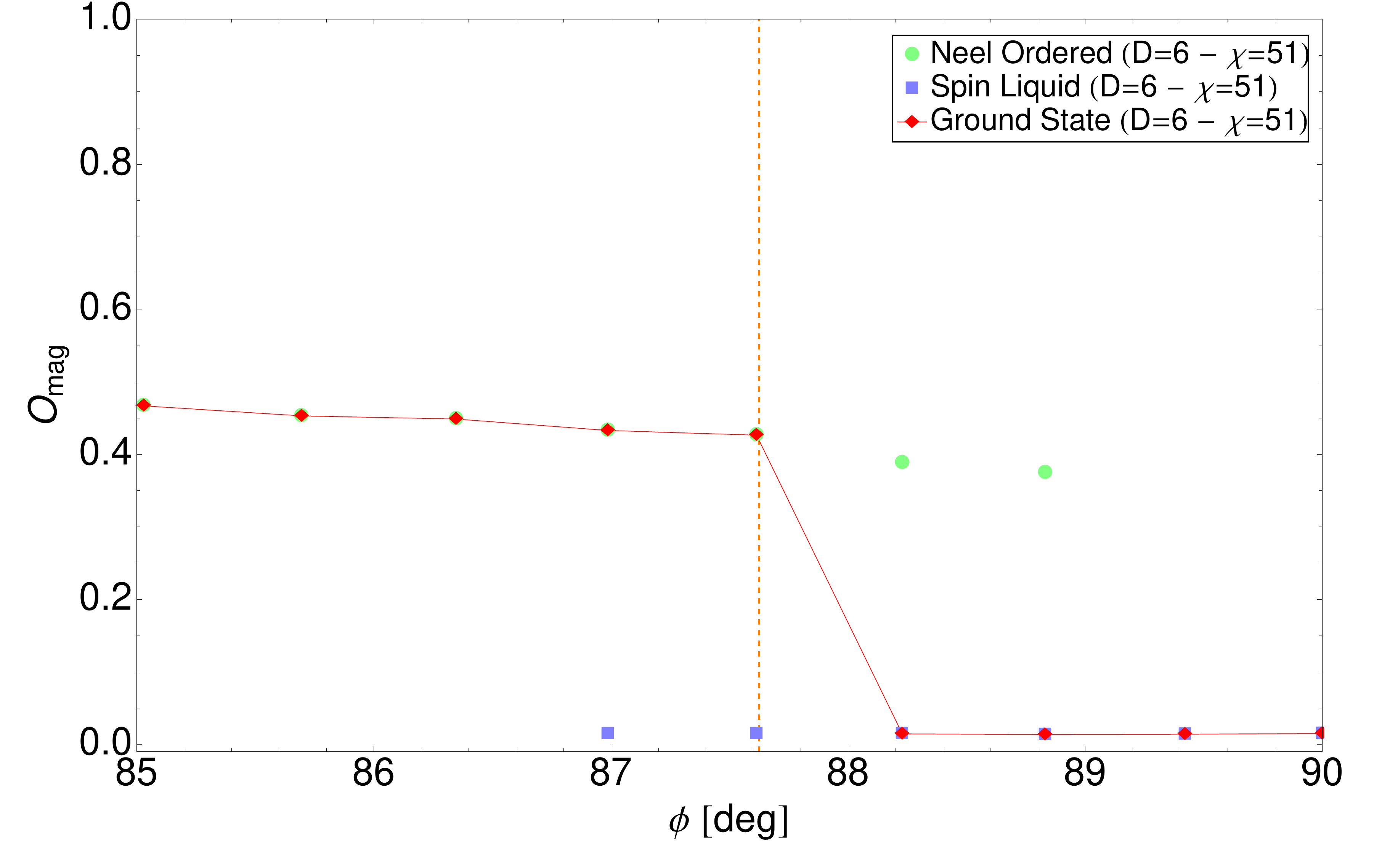} & \,\,\,\,\,\includegraphics[width=0.313\textwidth]{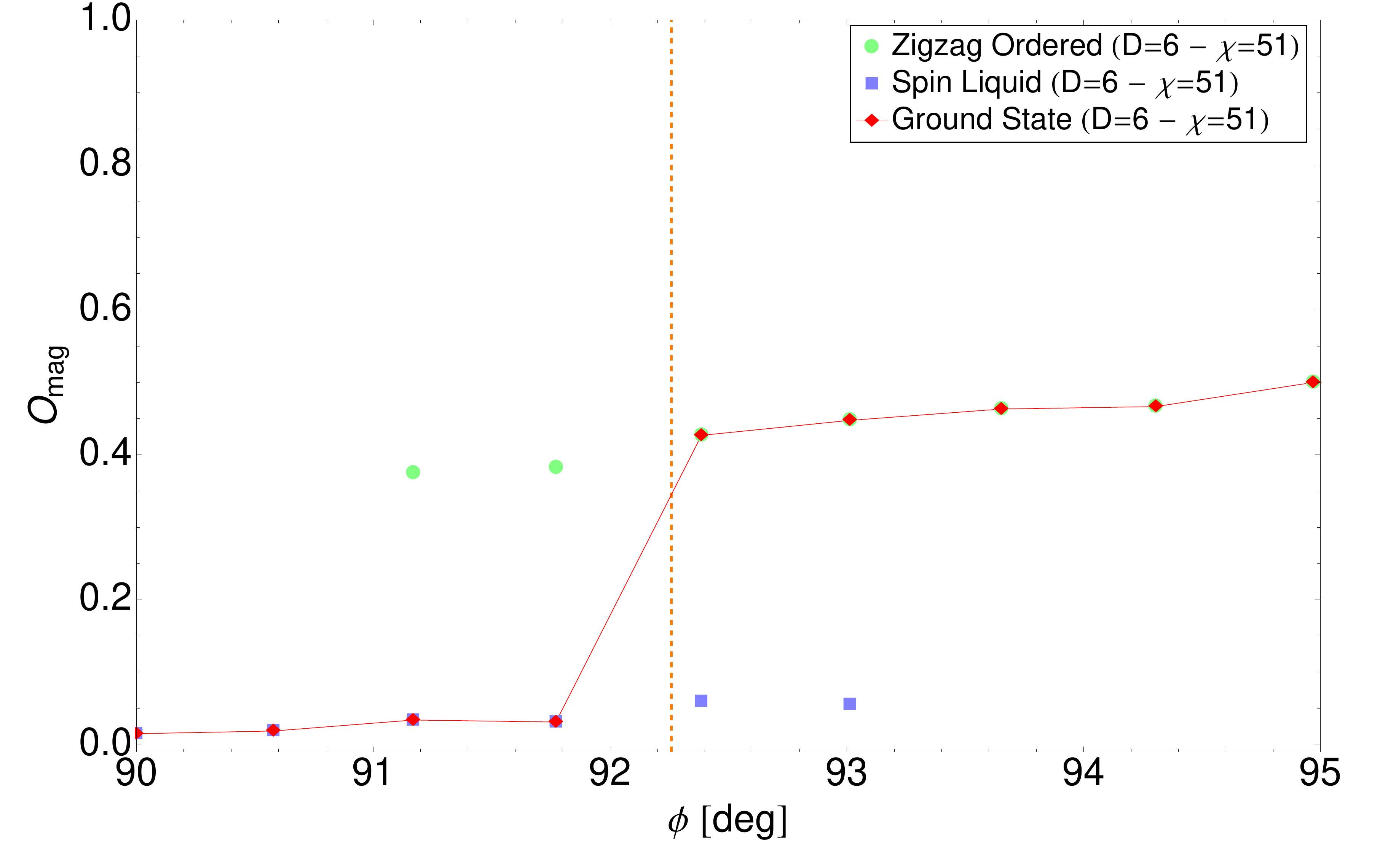} & \,\,\,\,\,\,\includegraphics[width=0.31\textwidth]{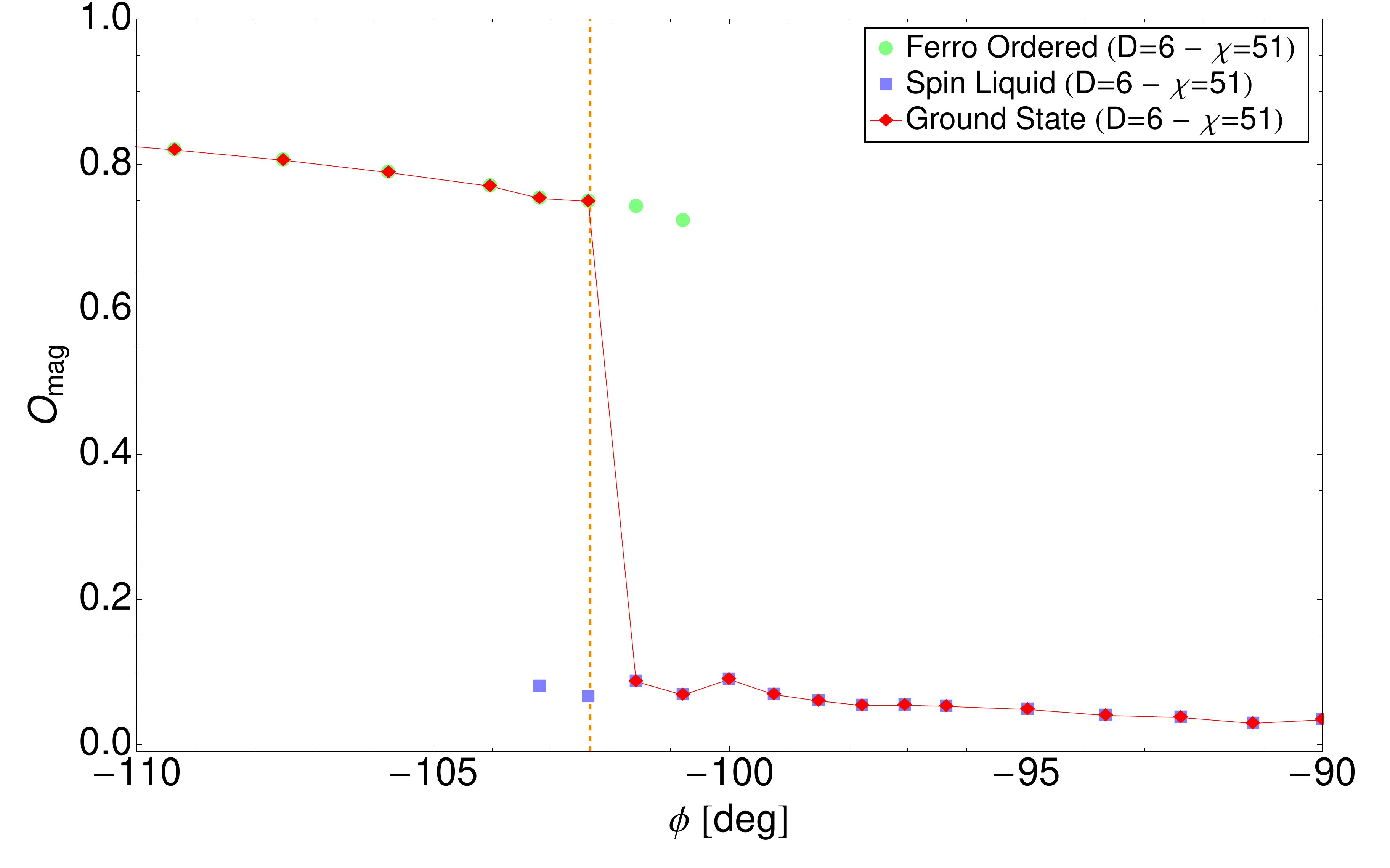} \\
 \end{tabular}
 \caption{ \label{fig:remaining_phases} Left: energy crossings (top) and $O_{mag}$ (bottom) for Néel and spin liquid phases. Center: energy crossings (top) and magnetization (bottom) for zigzag and spin liquid phases. Right: energy crossings (top) and magnetization (bottom) for ferromagnetic and spin liquid phases}
 \end{figure*}

\subsection{FSL-Ferromagnetic transition}
Keeping the sign of the couplings for the Kitaev term and flipping the interaction from AFM to FM for the Heisenberg term (this puts us in the third quadrant of the phase diagram, see Fig.~\ref{fig:order_patterns}) we find two phases: ferromagnetic and FSL, with the phase boundary being located at $\varphi \approx -102^{\circ}$ for $D = 6$ and a $D\rightarrow\infty$ extrapolated value of $\varphi_{\infty} \approx -106^{\circ}$. An energy crossing at a finite angle together with a discontinuity in the order parameters $O_{mag}$ and $O_{ferr}$ of comparable magnitude to that of the FSL-stripy transition again indicate that this is a first order phase transition, see Fig.~\ref{fig:remaining_phases} and the bottom plot in Fig.~\ref{fig:order_parameters}.

\subsection{ASL-Zigzag transition}
Switching the character of the interaction of the Kitaev term to AFM brings us to the second quadrant of the phase diagram where we find 3 phases: ferromagnetic, zigzag and ASL with the zigzag phase in between the ASL and ferromagnetic phases, see Fig.~\ref{fig:order_patterns}. The transition from ASL to zigzag is located at $\varphi \approx 92^{\circ}$ with $D = 6$ (a $D\rightarrow\infty$ extrapolation increases this value only very slightly to $\varphi_{\infty} \approx 93^{\circ}$), see Fig.~\ref{fig:remaining_phases}, whereas that from ferromagnetic to zigzag is found to be at $\varphi \approx 161^{\circ}$. These transitions also exhibit energy crossings at finite angles with the angle (strength of the transition) being significantly enhanced between the symmetry-broken phases. Discontinuities in the order parameters $O_{mag}$ $/$ $O_{zigzag}$ in the first case (see the top plot in Fig.~\ref{fig:order_parameters}) and $O_{ferro}$ $/$ $O_{zigzag}$ in the second (data not shown) again allow us to infer that these transitions are of first order type.

\subsection{ASL-Néel transition}
Finally, in the regime with all antiferromagnetic couplings $\varphi \in [0^{\circ},90^{\circ}]$, we find 2 phases: ASL and Néel. The phase transition is located at $\varphi \approx 88^{\circ}$ with $D = 6$ (extrapolating $D\rightarrow\infty$ again brings only a very slight lowering to $\varphi_{\infty} \approx 87^{\circ}$) and an energy crossing at a finite angle together with discontinuities in $O_{mag}$ $/$ $O_{\text{\emph{Néel}}}$ tell us that this transition is of first order type as well, see Fig.~\ref{fig:remaining_phases}. It should be said that the magnitude of the jumps in the ASL-Néel and ASL-zigzag cases is somewhat weaker than in the lower half of the phase diagram, see top plot in Fig.~\ref{fig:order_parameters}. \\

As noted previously, the regions corresponding to Néel and zigzag phases are connected via a 4-sublattice transformation.\cite{chaloupka2010kitaev} The same transformation connects the stripy phase to the ferromagnetic phase. As is well known, stronger quantum fluctuations in the Néel phase lead to a suppression in the order parameters compared to the ferromagnetic phase thus yielding weaker discontinuities in the phase transitions in the upper half of the phase diagram as compared to those found in the lower half. This is also nicely reflected in the order parameters shown in Fig.~\ref{fig:order_parameters} as their magnitudes at the phase transition points match quite well.\\
 
As a concluding remark, we note that a weak suppression in the order parameters as a function of bond dimension $D$ both for the spin liquid (as in Fig.~\ref{fig:energy_mag}) as well as the symmetry-broken phases (data not shown) is observed, albeit small so that ultimately we expect the discontinuities found to persist in the large $D$ limit. For a graphical summary of the results pertaining the span of the spin liquid phases see Figs. \ref{fig:order_patterns} and \ref{fig:order_parameters}.
 
\section{Summary and Discussion}

In summary, we have complemented previous studies \cite{schaffer2012quantum,jiang2011possible,chaloupka2010kitaev,chaloupka2013zigzag,he2013study} with accurate iPEPS simulations of the Kitaev-Heisenberg model in the thermodynamic limit. We have found excellent results when comparing with the pure Kitaev model even for moderate bond dimensions.\footnote{This shows that contrary to recent claims in Ref.~\onlinecite{he2013study}, the iPEPS ansatz can in fact properly encode the basic features of the Kitaev honeycomb model ground state.} The full ground-state phase diagram of the Kitaev-Heisenberg model was obtained where we were able to identify spin liquid phases covering regions of finite span.\\

In the case of the spin liquid with antiferromagnetic couplings we found first order phase transitions into symmetry-broken Néel and zigzag phases thus clarifying the nature of these transitions in the thermodynamic limit. In the case of the spin liquid with ferromagnetic couplings we have also found first order phase transitions leading to ferromagnetic and stripy ordered phases in contrast with results from previous small system studies \cite{jiang2011possible,chaloupka2010kitaev,chaloupka2013zigzag} and a reduced extent compared to these previous results. Given that the phase boundaries showed a systematic shift as a function of $D$ effectively increasing the size of the QSL phases we conclude that our $D=6$ results correspond to lower bounds. This means that upon increasing the bond dimension in our ansatz beyond $D=6$ we expect the FSL to span at least a region $\varphi \in [-102^{\circ},-80^{\circ}]$, whereas the ASL is expected to cover at least the region $\varphi \in [88^{\circ},92^{\circ}]$. The effect being more noticeable in the former case can be related to the nature of the competing phases (ferromagnetic and stripy) allowing the ferromagnetic spin liquid phase to profit more effectively from additional quantum fluctuations introduced as the bond dimension is increased.\\

Interesting future directions extending this study could involve a tensor network approach to the doped system, a direction which even at the mean field level has brought to light an interesting variety of superconducting phases arising in the model.\cite{you2012doping,hyart2012competition,okamoto2013global}

\begin{acknowledgements}

All simulations were performed using the Brutus cluster at ETH Zürich. Figures \ref{fig:energy_mag} and \ref{fig:correlators} were made using the LevelScheme \cite{levelscheme} package in Mathematica. This work is part of the D-ITP consortium, a program of the Netherlands Organisation for Scientific Research (NWO) that is funded by the Dutch Ministry of Education, Culture and Science (OCW). This work is also part of the FOR1807 Advanced Computational Methods for Strongly Correlated Quantum Systems research unit of the Deutsche Forschungsgemeinschaft (DFG). 

\end{acknowledgements}

\bibliography{hk_prb}

\end{document}